\RequirePackage{fix-cm}

\documentclass[twocolumn]{svjour3}          

\smartqed  
\usepackage{graphicx}
\usepackage{doi}
\usepackage[toc,page]{appendix}
\usepackage{authblk}
\usepackage[english]{babel}
\usepackage{amssymb}
\usepackage{hyperref}
\usepackage{amsmath}
\usepackage{multicol}
\usepackage{lineno}

\begin{document}

\title{Electroweak corrections to the angular coefficients in finite-$p_T$ $Z$-boson production and dilepton decay}



\author{\textbf{Rikkert Frederix, Timea Vitos} \endgraf  \bigskip \bigskip  \normalfont{LU-TP 20-42}
}

\institute{
              Theoretical Particle Physics, Department of Astronomy 
and Theoretical Physics, Lund University, S\"olvegatan\ 14\ A, SE-223\ 62 
Lund, Sweden \\
  \email{rikkert.frederix@thep.lu.se}  \\
               \email{timea.vitos@thep.lu.se}   (corresponding author)     
}

\date{\ }

\maketitle

\begin{abstract}
We present next-to-leading order (NLO) electroweak corrections
to the dominant five angular coefficients parametrizing the Drell-Yan
process in the $Z$-boson mass peak range for finite-$p_T$ vector boson
production.  The results are presented differentially in the vector
boson transverse momentum. The Lam-Tung violating difference $A_0-A_2$
is examined alongside the coefficients. A single lepton transverse
momentum cut is needed in the case of electroweak corrections to avoid
a double singularity in the photon induced diagrams, and the
dependence on the value of this cut is examined. We compare the
electroweak corrections to the angular coefficients to the NLO QCD
corrections, including the single lepton cut. The size of the single
lepton cut is found to affect the two coefficients $A_0$ and
$A_2$ to largest extent. The relative size of the electroweak
corrections to the coefficients  is moderate
for all single lepton cut values, and by extrapolation to the
inclusive results, is moderate also for the full dilepton phase space
case. However, for the Lam-Tung violation, there is a significant contribution from the electroweak corrections for low $p_T$ of the lepton pair.     
\end{abstract}

\section{Introduction} 
\label{intro}
With a new era of LHC runs lying ahead, accounting for Standard Model background signatures with great accuracy becomes increasingly important. Electroweak (EW) corrections are, by nature of the magnitudes of the gauge couplings in the Standard Model, at energy scales relevant to present collisions, an order smaller than the strong corrections. This implies that EW corrections are, when accuracy is difficult to obtain, not of primary interest. However, in processes where precision reaches that of predictions at next-to-next-to-leading order (NNLO) QCD, an inclusion of electroweak corrections is necessary. 

The Drell-Yan process of lepton pair production in hadron-hadron collisions has been of significant interest for the past years in particle physics because of its high availability in experiment and important implications for the parton model \cite{Peng:2014hta}. Due to the clear signatures in experiment, this process together with deep inelastic scattering, are the benchmark processes for determination of parton distribution functions. A precise theory prediction for the Drell-Yan process is of high importance for fundamental particle physics research. The related process of charged heavy vector boson production and decay, with a final state  $l^{\pm}{\nu_l}$ requires similar analysis, but due to difficulties in missing energy measurements, the signature for this process is less propitious than for the neutral current process. 

The lepton pair production in hadron-hadron collisions was first discussed in Ref.~\cite{Drell:1970wh}. The Drell-Yan process was examined in the parton model, and the leading order $\mathcal{O}(\alpha^2)$ expression for the process in terms of parton distribution functions was presented. This was shortly followed by NLO QCD corrections to the process in Ref.~\cite{Altarelli:1978id}, obtaining large $\mathcal{O}(\alpha^2\alpha_S)$ corrections. During the following years, one benchmark work was that of Ref.~\cite{Collins:1977iv} where the angular distribution of the lepton pair was investigated. In the work \cite{Lam:1978pu,Lam:1980uc} the cross section in terms of structure functions for the hadronic current was studied, assuming solely a (virtual) photon interaction and introducing the Lam-Tung relation, in analogy with the Callan-Gross relation for deep inelastic scattering \cite{Callan:1969uq}. Dilepton production arising from both virtual photon and $Z$-boson decays was first covered in Ref.~\cite{Chaichian:1981va}, where the first five of the angular coefficients were considered at finite-$p_T$ vector boson production. The remaining three angular coefficients vanish at order $\mathcal{O}(\alpha^2\alpha_S)$ and hence are not considered up to NNLO QCD.

In contrast to the zero-$p_T$ Drell-Yan process  $p p \rightarrow Z \rightarrow l^+ l^-$, in this work we consider the finite-$p_T$ Drell-Yan process $p p \rightarrow Z +X \rightarrow l^+ l^- +X$, whose leading order is given by $\mathcal{O}(\alpha^2\alpha_S)$. The kinematics of this process is parametrized by eight angular coefficients \cite{Collins:1977iv}, each containing information of the spin state of the vector boson. Calculation of these angular coefficients has been presented previously up to NNLO QCD in Ref.~\cite{Mirkes:1994dp,Gauld:2017tww}. Two of the coefficients, $A_0$ and $A_2$ satisfy at leading order $\mathcal{O}(\alpha^2\alpha_S)$ the Lam-Tung relation $A_0-A_2=0$ \cite{Lam:1978pu,Lam:1980uc}, a manifestation of the spin properties of the vector boson and the leptons. Measurements of these angular coefficients have been performed previously at Tevatron \cite{Aaltonen:2010zza,Abazov:2007jy}, at CMS \cite{Khachatryan:2015paa} and most recently at ATLAS \cite{Aad:2016izn}. The experimental data shows a larger violation of the Lam-Tung relation than is predicted at NNLO QCD $\mathcal{O}(\alpha^2\alpha_S^3)$ in Ref.~\cite{Gauld:2017tww}. Efforts to describe this discrepancy have been made \cite{Peng:2015spa} in terms of non-perturbative effects \cite{Brandenburg:1994wf} and spin asymmetries \cite{Boer:1999mm}. The motivation of the present work is to investigate the electroweak effects at fixed order to this process.

The outline of the article is the following. In Sec.~\ref{section:theory}, we present the theoretical setup and address electroweak corrections and how these are to be treated in the prevailing work. In Sec.~\ref{section:setup}, we discuss the numerical setup, the selection criteria, and discuss our treatment of the theoretical uncertainties. In Sec.~\ref{section:results} we present the results and finally discuss these in Sec.~\ref{section:conclusions}.

\section{Theoretical setup}
\label{section:theory}

\subsection{Angular coefficients}

To introduce notation, we consider the following high energy proton-proton ($pp$) collisions to a dilepton final state ($l^+l^-$),
\begin{eqnarray}
p(k_1) + p(k_2) \rightarrow  l^+(k_3) + l^-(k_4) + X(k_5) 
\end{eqnarray}
where we indicate the momentum of each particle in brackets and $X$ has been introduced as the recoil to the lepton pair. In the region where the lepton pair invariant mass $m_{ll}^2=(k_3+k_4)^2$ is in the $Z$-boson pole range, the dominant contribution to the process is the $Z$-boson production and decay
\begin{eqnarray}\label{eq:process}
p+p\rightarrow Z + X \rightarrow l^+ +l^- + X,
\end{eqnarray}
while the photon mediated process is present but subdominant due to the large virtuality of the photon. The $Z$-boson momentum is then given by the lepton pair momentum, $p_Z=k_3+k_4$ with transverse momentum $p_{T,Z}$ and rapidity $y_Z$. The spin polarization of the vector boson directly affects the angular distribution of the lepton pair. For the zero-$p_T$ Drell-Yan process this yields a $(1+\cos \theta)^2$ dependence, in similarity to the $W$-boson production, however, at finite-$p_T$ this simple dependence changes \cite{Mirkes:1992hu}. 

The differential cross section for the process can be expanded in
terms of real spherical harmonics and associated coefficients which
bare the dependencies on the vector boson kinematics. The angular
coefficients appearing in the expansion are ratios between the
different spin states and the unpolarized cross section, and can be
analytically determined at next-to-leading order in QCD
\cite{Mirkes:1992hu,Mirkes:1994dp}. Numerous notations exist for the
decomposition of the differential cross section into structure
functions or helicity amplitudes. We follow the one in
Ref.~\cite{Gauld:2017tww,Mirkes:1992hu}, where the decomposition is
into eight frame-dependent angular coefficients denoted by $A_i$ with
$i=0,\ldots,7$. In this work the choice of frame is the Collins-Soper
frame (see below), in which case the (negatively charged) lepton
angular coordinates in the frame are $\phi$ (azimuthal) and $\theta$
(polar). The expansion in this notation reads
\begin{eqnarray}\label{eq:decomposition}
\begin{split}
\frac{\text{d}\sigma}{\text{d}p_{T,Z}\text{d}y_Z\text{d}m_{ll}\text{d}\Omega}=&\frac{3}{16\pi}\frac{\text{d}\sigma^{U+L}}{\text{d}p_{T,Z}\text{d}y_Z\text{d}m_{ll}}\\
& \bigg((1+\cos^2\theta)+A_0\frac{1}{2}(1-3\cos^2 \theta)\\
 &   +A_1 \sin 2\theta \cos \phi \\
& + A_2 \frac{1}{2}\sin^2\theta \cos 2\phi \\
& +A_3\sin \theta \cos \phi +A_4 \cos \theta \\
 & +A_5 \sin^2 \theta \sin2\phi+A_6 \sin2\theta \sin\phi \\
 &+A_7 \sin \theta \sin \phi\bigg),
\end{split}
\end{eqnarray}
where $\sigma^{U+L}$ denotes the unpolarized cross section. 

At leading order, the coefficients are (linear combinations of) the structure functions of the hadronic tensor $W_{\mu \nu}$ in the amplitude decomposition $\mathcal{M}\propto W_{\mu \nu}L^{\mu \nu}$ with the probing leptonic tensor $L^{\mu \nu}$. At order $\mathcal{O}(\alpha^2\alpha_S)$, the Lam-Tung relation between two of these coefficients reads 
\begin{eqnarray}\label{eq:LT}
A_0-A_2=0, 
\end{eqnarray}
which can be derived using the properties of the amplitudes.

The real spherical harmonics $Y_{lm}(\theta,\phi)$ present in
Eq.~(\ref{eq:decomposition}) are of order $l \le 2$. Following
Ref.~\cite{Gauld:2017tww}, we use the orthogonality relation
\begin{eqnarray}\label{eq:orthogonality}
\int Y_{lm}(\theta,\phi) Y_{l'm'}(\theta,\phi) \text{d}\Omega=\delta_{ll'}\delta_{mm'}
\end{eqnarray}
to project out the angular coefficients $A_i(m_{ll},p_{T,Z},y_Z)$,
using a weighted normalization
\begin{eqnarray}\label{eq:ratio_def}
\langle f(\theta,\phi)\rangle=\frac{\int \text{d}\Omega \text{d}\sigma f(\theta,\phi)}{\int \text{d}\Omega \text{d}\sigma},
\end{eqnarray}
with the usual  solid angle differential
$\text{d}\Omega=\text{d}\phi \text{d}\cos\theta$. This can be
implemented in Monte Carlo calculation by reweighting each event with
the corresponding function $f(\theta,\phi)$.

A note on the chosen frame of reference: a suitable choice is a rest
frame of the vector boson, in which the angular dependence of the
final state leptons can be analyzed. We perform the calculations in
the Collins-Soper reference frame~\cite{Collins:1977iv}, in which
previous works on the angular coefficients have been performed, and
the frame adopted by LHC measurements. This frame is defined as the
rest frame of the heavy vector boson in which the $z$-axis is chosen
to be along the external bisector of the two incoming parton momenta,
with the positive direction in the same direction as the lepton pair
in the laboratory frame.  The $x$-axis is chosen to be along the
bisector of the incoming parton momenta, with the positive direction
opposite to the sum of the two incoming parton momenta. The $y$-axis is
then chosen to complete a right-handed Cartesian coordinate system.

\subsection{NLO electroweak corrections}
The NLO EW corrections to the dilepton+jet final state have been first
computed by Denner \textit{et al.}~\cite{Denner:2011vu}. We classify
the contributions to the perturbative structure of the cross section
in Eq.~(\ref{eq:process}) up to next-to-leading order in the gauge
couplings in the following manner
\begin{align}
\text{d}\sigma^{\text{(LO)}}&= \alpha^2\alpha_S B_1,\\
\text{d}\sigma^{\text{(NLO QCD)}}& = \alpha^2\alpha_S B_1 + \alpha^2\alpha_S^2  C_1, \label{eq:nloqcd}\\
\text{d}\sigma^{\text{(NLO EW)}}&= \alpha^2\alpha_S B_1+\alpha^3B_2+\alpha^3\alpha_S C_2, \label{eq:nloew}
\end{align} 
where the $B_i$ (at LO) and $C_i$ (at NLO) label finite values
obtained by evaluation of the corresponding Born and virtual and real
emission diagrams, respectively. In the NLO EW term, we also include
the subleading term at Born-level, $\alpha^3B_2$. Note that the last term in Eq. (\ref{eq:nloew}) is to be interpreted as both electroweak $\mathcal{O}(\alpha)$ corrections to the LO term $\alpha^2\alpha_SB_1$, but also as the QCD corrections at $\mathcal{O}(\alpha_S)$ to the term $\alpha^3B_2$. By not including the $\alpha^3B_2$ term at LO in Eq. (\ref{eq:nloqcd}), this term is also omitted in the NLO QCD corrections (which is consistent, as this contribution is negligible to the $\alpha^2\alpha_S^2C_1$ term). We consistently include the photon induced processes in order to obtain
the correct IR cancellations. We obtain the cross section up to NLO
QCD+EW using the additive approach,

\begin{eqnarray}
\begin{split}
\text{d}\sigma^{\text{(NLO QCD+EW)}}=\ & \text{d}\sigma^{\text{(NLO
    QCD)}}+\text{d}\sigma^{\text{(NLO
    EW)}}\\
&  -\text{d}\sigma^{\text{(LO)}}.
\end{split}
\end{eqnarray}  

For the electroweak corrections to the angular coefficients, in the presence of real photon
emission from the external leptons, the expansion in Eq.~(\ref{eq:decomposition}) is a priori not valid. The three-body decay $Z
\rightarrow l^+ l^- \gamma$ alters the kinematics. As a direct
distinction of such a hard photon is not possible, we attempt to
analyze the angular coefficients as given by this expansion and examine to what extent this expansion
is valid in the case of electroweak corrections. Hence, a direct
comparison to the theoretically derived angular coefficients would not
be well-motivated. We perform a comparison between the QCD corrections
and the QCD+EW corrections to the coefficients as obtained by the
projection Eq.~(\ref{eq:ratio_def}) to qualitatively examine the
effect.

\section{Numerical setup}\label{section:setup}

\subsection{Basic cuts and parameters}

For the evaluation of the differential cross sections, we use 
\textsc{MadGraph5\_aMC@NLO}~\cite{Alwall:2014hca,Frederix:2018nkq} for the process $p p
\rightarrow l^+ l^- j$ at $\sqrt{s}=8$ TeV at fixed order. We work in
a five-flavor scheme, where all lepton and quark masses except for
the top quark are set to zero. In order to generate the lepton pair at
non-zero transverse momentum, we add the parton $j$ to the process,
which can be a (anti) quark, gluon or photon\footnote{The process where the recoil is a heavy vector boson is omitted from this work, albeit being of the order of interest. See Ref.~\cite{1810.11034} for discussion of the angular coefficients in such cases.}.

For the input parameters, we adapt the
complex-mass-scheme~\cite{Denner:1999gp,Denner:2005fg}, in which (in
our case) the masses of the heavy vector bosons and the top quark are
treated as complex numbers, thus rendering the dependent parameters
complex. In order to maintain correct cancellations in the subtraction
schemes, we use the $\bar{G}_{\mu}$ scheme~\cite{Frederix:2018nkq}, in
which the $G_{\mu}$ constant obtains a phase, compensating for the
phase of the electroweak coupling constant $\alpha$. The input masses
and widths of the relevant particles which are used in the assignment
of complex masses are
\begin{eqnarray}
\begin{split}
  G_{\mu}&=1.16639 \times 10^{-5} \text{ GeV}^{-2}, \\
 m_Z&=91.154 \text{ GeV}, \qquad \Gamma_Z=2.4956 \text{ GeV}    \\
 m_W&=80.358 \text{ GeV},\qquad \Gamma_W=2.0890 \text{ GeV}, \\
m_t&=173.34 \text{ GeV},\qquad \Gamma_t=1.36918 \text{ GeV}.
\end{split}
\end{eqnarray}
The two parity-odd coefficients $A_3$ and $A_4$ show high sensitivity
to the value of the weak mixing angle
$\theta_W$~\cite{Aad:2016izn}. This is remedied by including the
one-loop correction to the $\rho$-parameter in the LO and the NLO QCD
predictions. Using the complex masses for the particles, this gives
the effective value of
\begin{eqnarray}
\sin^2 \theta_W=1-\left(\frac{\mu_W}{\mu_Z}\right)^2+\Delta\rho\left(\frac{\mu_W}{\mu_Z}\right)^2,
\end{eqnarray}
with the one-loop correction included from Ref.~\cite{Fleischer:1993ub}
\begin{eqnarray}\label{eq:rho}
\Delta\rho=\frac{\sqrt{2}\bar{G}_{\mu}}{16\pi^2}3\mu_t^2,
\end{eqnarray}
and applied with the complex-valued $\bar{G}_\mu$. The $\mu_Z$,
$\mu_W$ and $\mu_t$ are the complex masses for the $Z$- and $W$-bosons and the top quark, respectively. In the NLO EW predictions, these loop effects are already included as part of the loop corrections in the
electroweak sector. The inclusion of the $\Delta\rho$ correction at this order in the electroweak corrections is redundant: a consistent inclusion of it is present in the loop diagrams. However, for a more elaborate and comprehensive treatment of the input parameters, one may employ the effective $\sin^2\theta_{eff}^l$ scheme ~\cite{Chiesa:2019nqb}, in which a different set of input parameters are used than to the usual $G_{\mu}$ scheme, and the subtraction of the double counting in the electroweak correction is performed. For the purpose of the present work, our inclusion of the electroweak effects in the electroweak mixing angle is consistent and sufficient.  

Photon recombination is performed with all light charged fermions on
equal footing. A fermion (lepton or quark) is recombined with a photon
if the distance in the $\eta-\phi$ plane,
$R=\sqrt{\Delta\eta^2+\Delta\phi^2}$, fulfills $R<0.1$. Upon
recombination, the photon momentum is added to the fermion momentum,
and the former is removed from the list of external particles.

After recombination, the following basic cuts for event selection are
applied. We use the narrow cut on the invariant mass of the lepton
pair, the same as is used in the ATLAS measurement: $m_{ll} \in
[80,100]$ GeV. This we do in order for the $Z$-boson diagrams to be
the dominant contribution, allowing for a determination of also the
parity-odd coefficients $A_3$ and $A_4$. Events with lepton pair
transverse momentum $p_{T,Z}>11.4$ GeV are selected and results
presented in the $p_{T,Z}$ range of [11.4,400] GeV. We use no cuts on
the jet transverse momentum and demand no reconstructed jet in the
final state.

\begin{figure}[htb!]
\begin{center}
\includegraphics[scale=0.4]{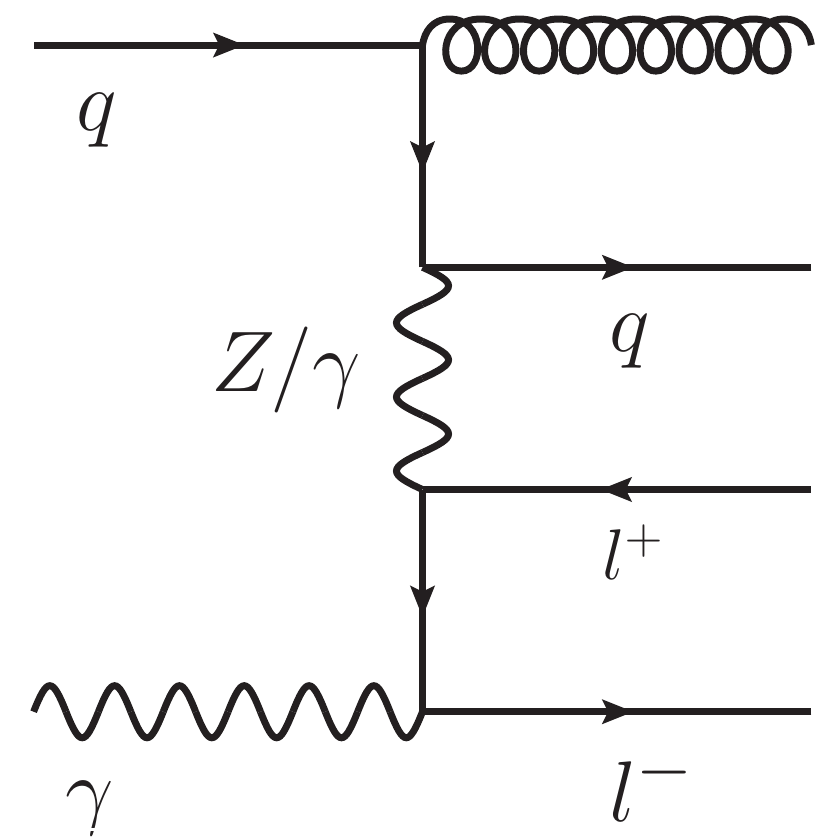}
\caption{An illustrative Feynman diagram from the photon induced real
  emission electroweak corrections leading to a double singularity
  uncanceled by virtual diagrams at the same order.}
\label{fig:div}
\end{center}
\end{figure}

The soft and collinear divergences in the real-emission phase-space
integration are canceled with the virtual corrections according to the
Kinoshita-Lee-Nauenberg theorem
\cite{Kinoshita:1962ur,Lee:1964is}. For (N)NLO corrections in QCD, the
requirement of non-zero $p_T$ for the lepton pair (together with their
invariant mass cut) is enough to render the process finite. However,
when EW corrections are involved these cuts are not enough: in the
photon induced, real emission diagrams, of which an illustrative
example can be seen in Fig.~\ref{fig:div}, there can be double
soft/collinear divergences. In the example diagram, the internal quark
and lepton propagators can go on-shell simultaneously if the gluon and
the electron are both collinear to the incoming partons. Such a double
singularity is not canceled with a loop diagram at this order. In
order to avoid these divergences, we place an additional single lepton
transverse momentum cut (on both the negatively and positively charged
leptons) for obtaining the differential distributions. This cut alters
the setup of the ATLAS and CMS measurements, which are performed
inclusively in the lepton transverse momentum. Moreover, as this cut
no longer allows for a full phase space inclusion of the final state
leptons, the orthogonality relation, Eq.~(\ref{eq:orthogonality}),
which we use to compute the angular coefficients is, strictly
speaking, no longer valid.  To examine to which extent this cut
affects the result, we present the differential distributions for
three different values of the lepton transverse momentum cut: for
$p_{T,l}>\{2.0,5.0,8.0\}$ GeV. Thus, we do not expect to be able to
directly compare our predictions to data (which we therefore also do
not show), but we should be able to address the size of the NLO EW
corrections, at least in a qualitative manner, since both the LO and
the NLO predictions will be affected by the cut. 

The double singularity appearing in the real emission diagrams can be avoided by introducing the finite masses of the leptons. While this approach would avoid the singularities, the small masses of the electrons would yield logarithmically enhanced contributions, arising from a large difference between the scale and the mass. In order to avoid this issue, in this work we implement the technical transverse momentum cut on the single lepton and see to which extent this affects the final result.

\subsection{Scale and PDF}
For the scale choice, we follow Ref.~\cite{Gauld:2017tww} for the central value and the variation. We perform an uncorrelated variation of the renormalization and factorization scales in the numerator and denominator in Eq.~(\ref{eq:ratio_def}) of the angular coefficients. Note that also the coefficient $A_0$ needs to be brought to a single quotient expression in order to apply this uncorrelated scale variation. For the choice of the central value, the transverse energy of the lepton pair is used,
\begin{eqnarray}
\mu_0=\sqrt{m_{ll}^2+p_{T,Z}^2}.
\end{eqnarray}
Independently, we perform a 9-point scale variation for the numerator and denominator in each case, varying between $\frac{1}{2} \le \mu^{\text{num,den}}_{\text{R,F}}/\mu_0 \le 2$, and combining them in a way that $\frac{1}{2}\le \mu^{\text{num}}_{\text{R,F}}/\mu^{\text{den}}_{\text{R,F}}\le 2$ holds. The envelope is taken to be among these 31 possible combinations. The statistical error of the ratios is calculated by the usual propagation of errors.

For the numerical calculations, we use the PDF set
\textsc{LUXqed17\_plus\_PDF4LHC15\_nnlo\_100}
~\cite{Manohar:2017eqh} from the \textsc{LHAPDF6}
library~\cite{Buckley:2014ana}, including the photon content in a more
robust way, at all orders of the calculation. A comparison to results
obtained with the NNPDF2.3 set~\cite{Ball:2012cx} has been made, which
is a set with a larger photon luminosity (and with much larger
uncertainties), but the difference in the central values is negligible
for the observables we consider in the following. The PDF
uncertainties enter in the same manner in the numerator and
denominator of Eq.~(\ref{eq:ratio_def}), thus canceling their effects
in the angular coefficients to a large extent. This is the reason that
they are omitted in this work.

\section{Results}
\label{section:results}

\begin{figure*}[htb!]
\begin{multicols}{2}
    \includegraphics[width=0.9\linewidth]{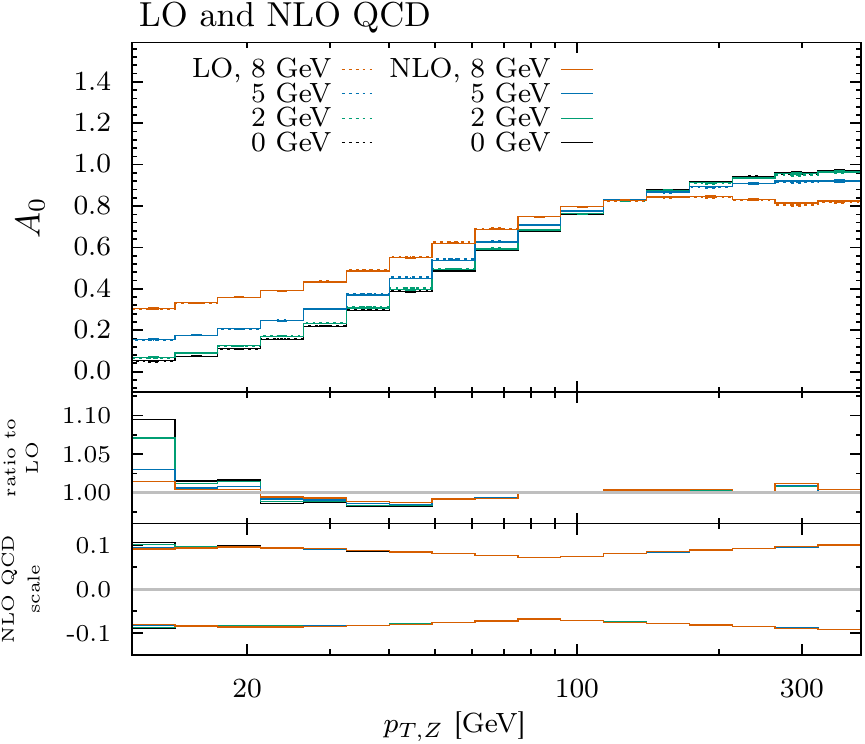}\par 
    \includegraphics[width=0.9\linewidth]{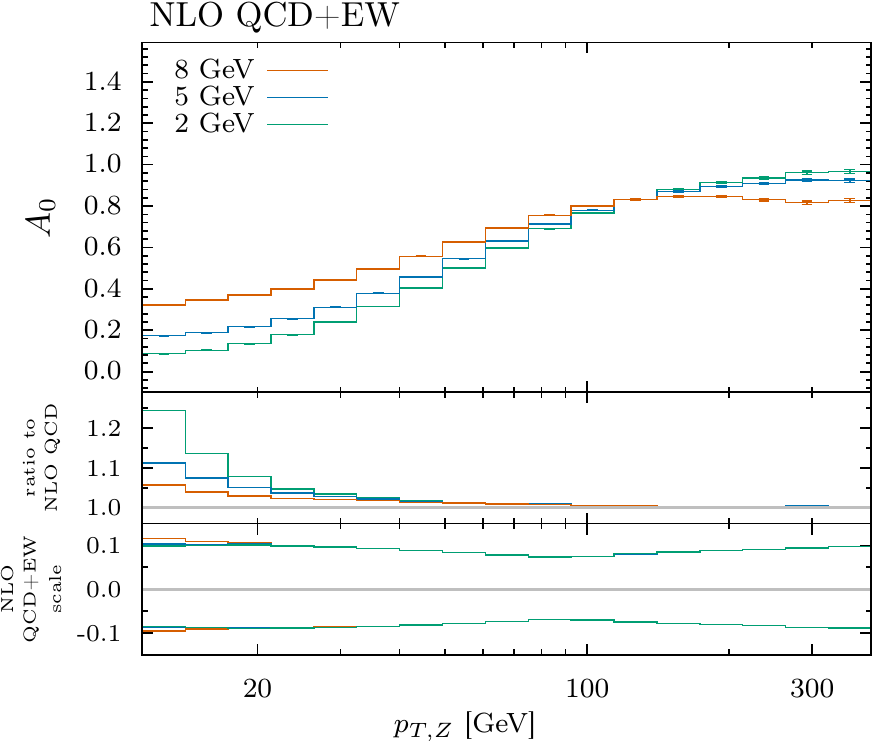}\par 
    \end{multicols}
\begin{multicols}{2}
    \includegraphics[width=0.9\linewidth]{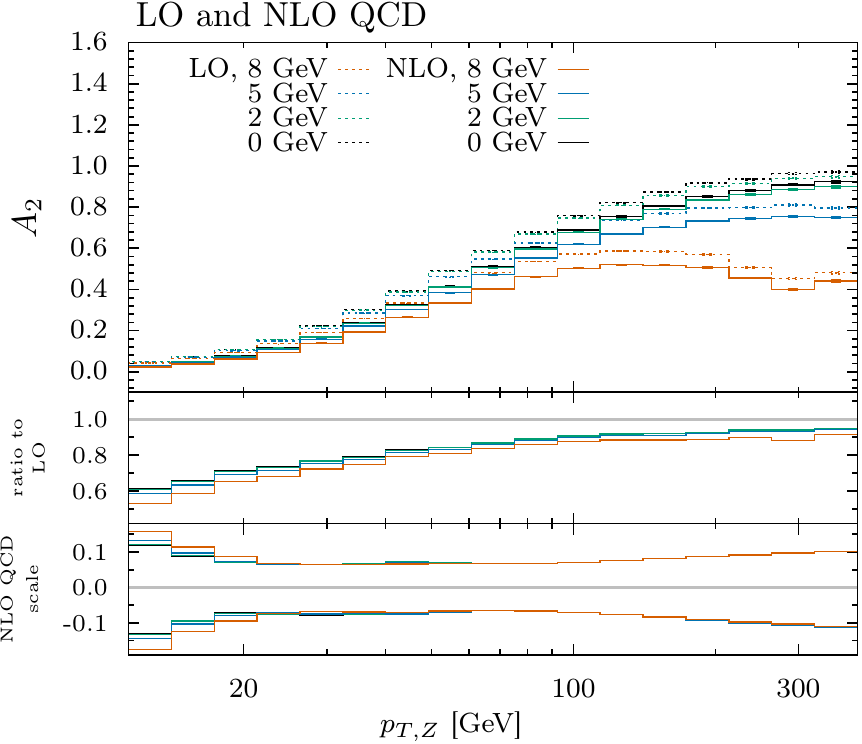}\par 
    \includegraphics[width=0.9\linewidth]{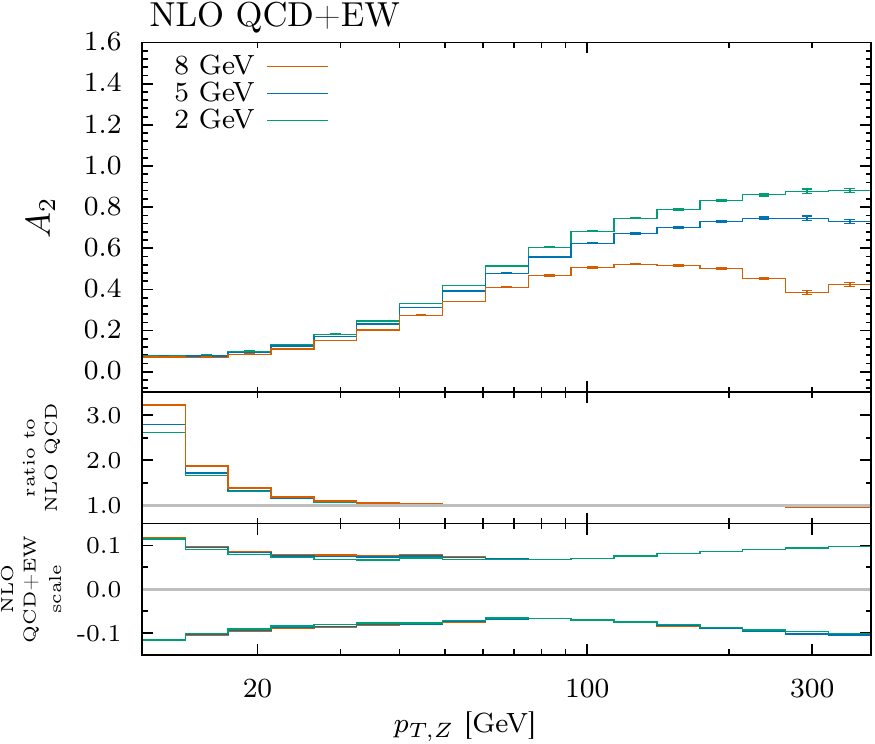}\par
\end{multicols}
\caption{The two largest angular coefficients $A_0$ (top) and $A_2$ (bottom), the LO and NLO QCD (left) and the NLO QCD+EW contribution (right) with corresponding ratios and scale uncertainties. The four (left) and three (right) different curves indicate the result with different values of the single lepton $p_T$ cut. See text for more details.}
\label{fig:A0_A2}
\end{figure*}

\begin{figure*}[htb!]
\begin{multicols}{2}
    \includegraphics[width=0.9\linewidth]{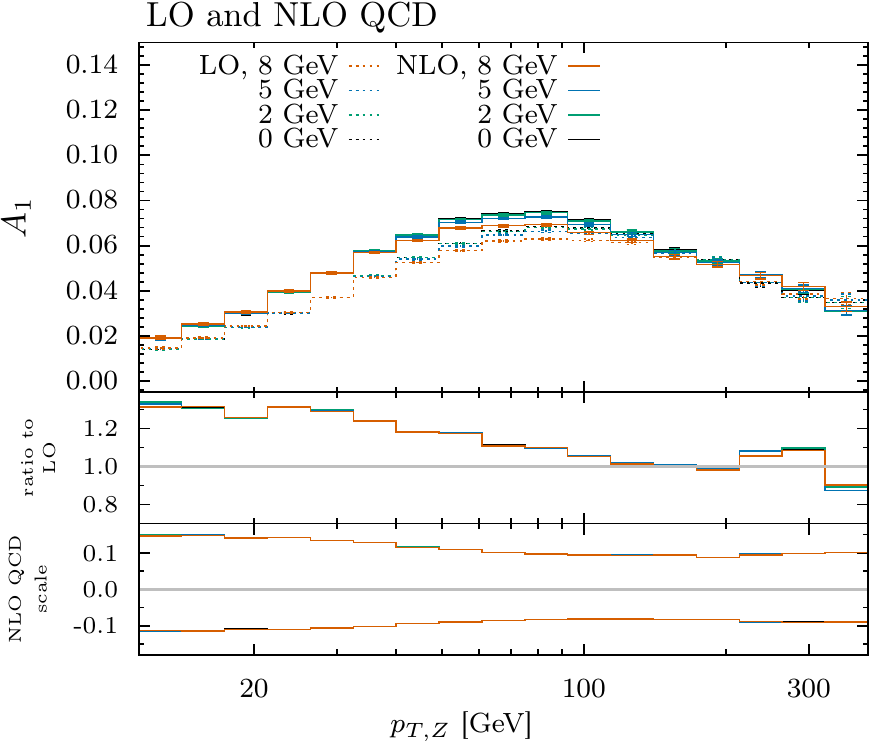}\par 
    \includegraphics[width=0.9\linewidth]{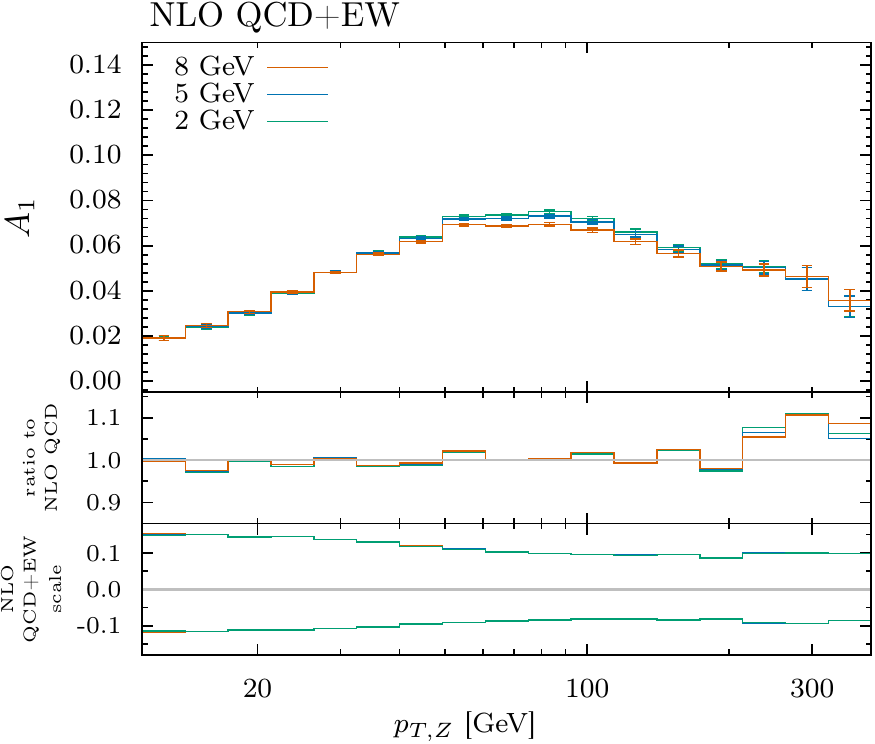}\par 
    \end{multicols}
\begin{multicols}{2}
    \includegraphics[width=0.9\linewidth]{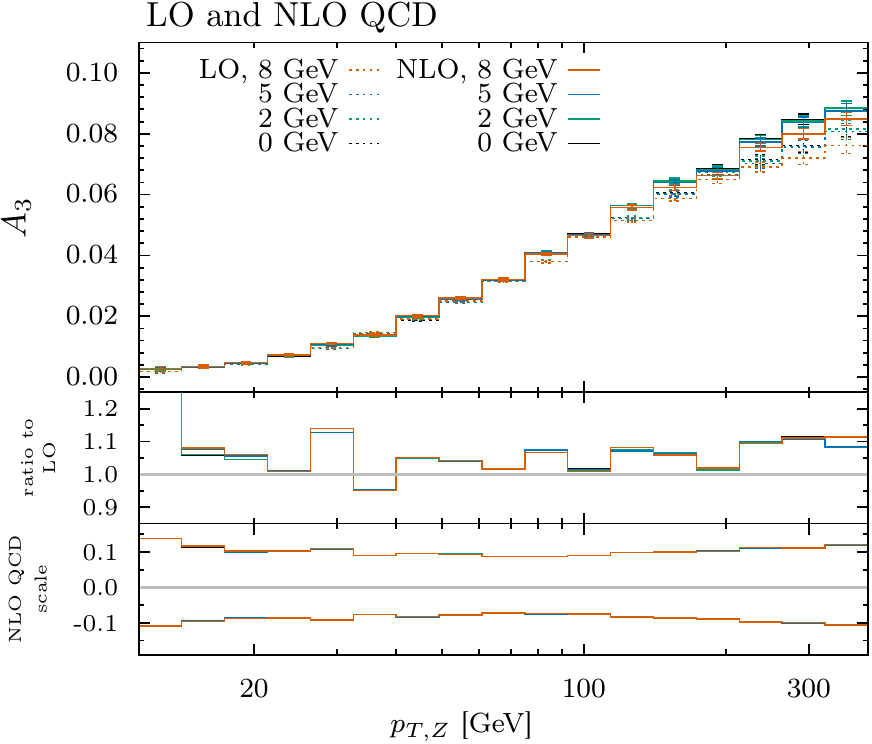}\par 
    \includegraphics[width=0.9\linewidth]{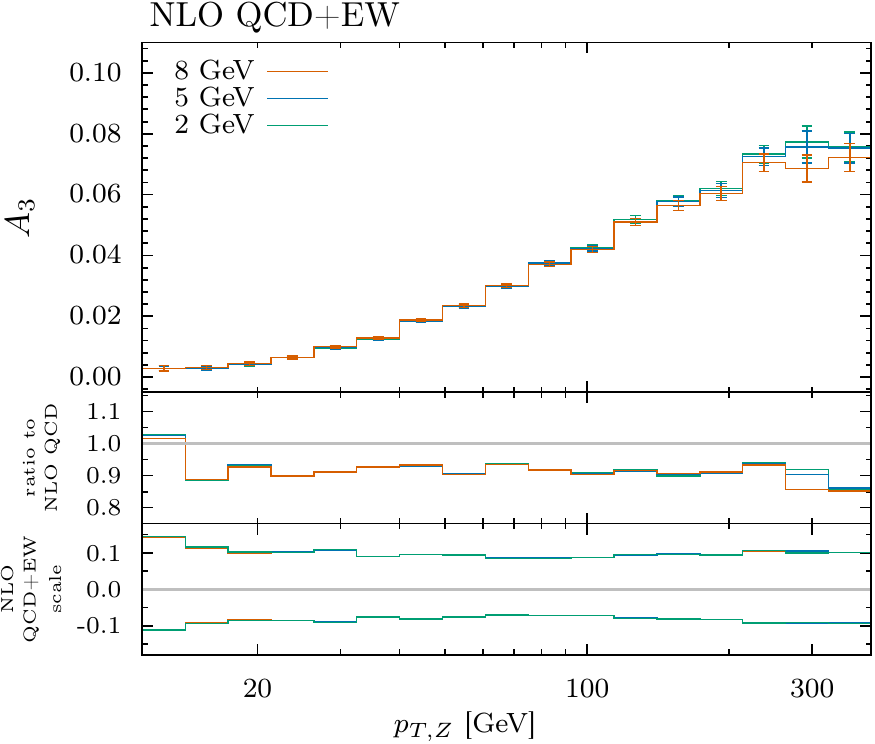}\par
\end{multicols}
\caption{The two angular coefficients $A_1$ (top) and $A_3$ (bottom),
  the LO and NLO QCD (left) and the NLO QCD+EW contribution (right)
  with corresponding ratios and scale uncertainties. The four (left)
  and three (right) different curves indicate the result with
  different values of the single lepton $p_T$ cut. See text for more
  details.}
\label{fig:A1_A3}
\end{figure*}

\begin{figure*}
\begin{multicols}{2}
    \includegraphics[width=0.9\linewidth]{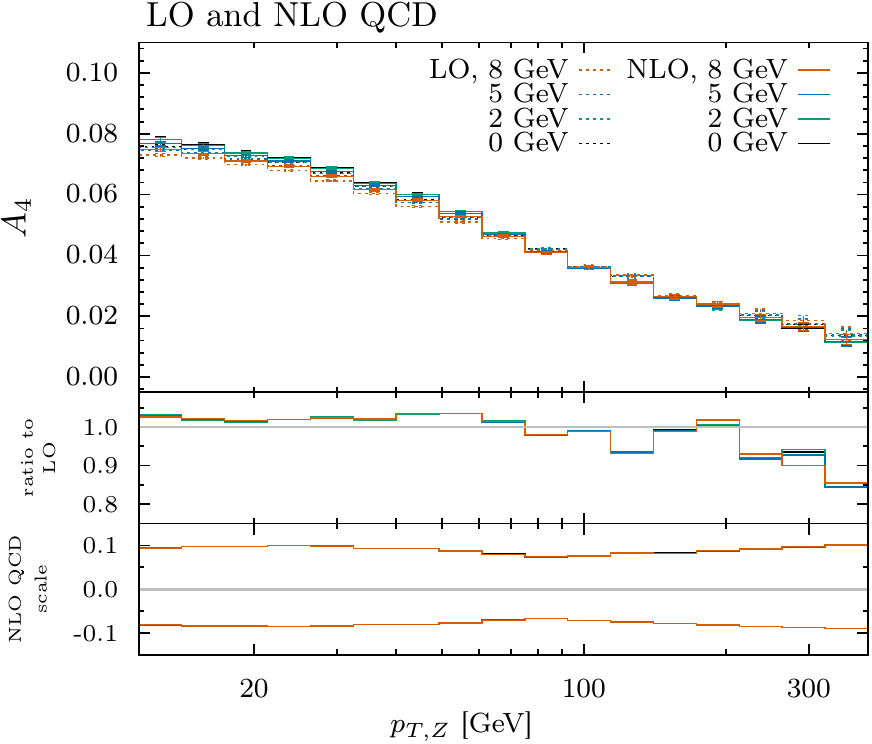}\par 
    \includegraphics[width=0.9\linewidth]{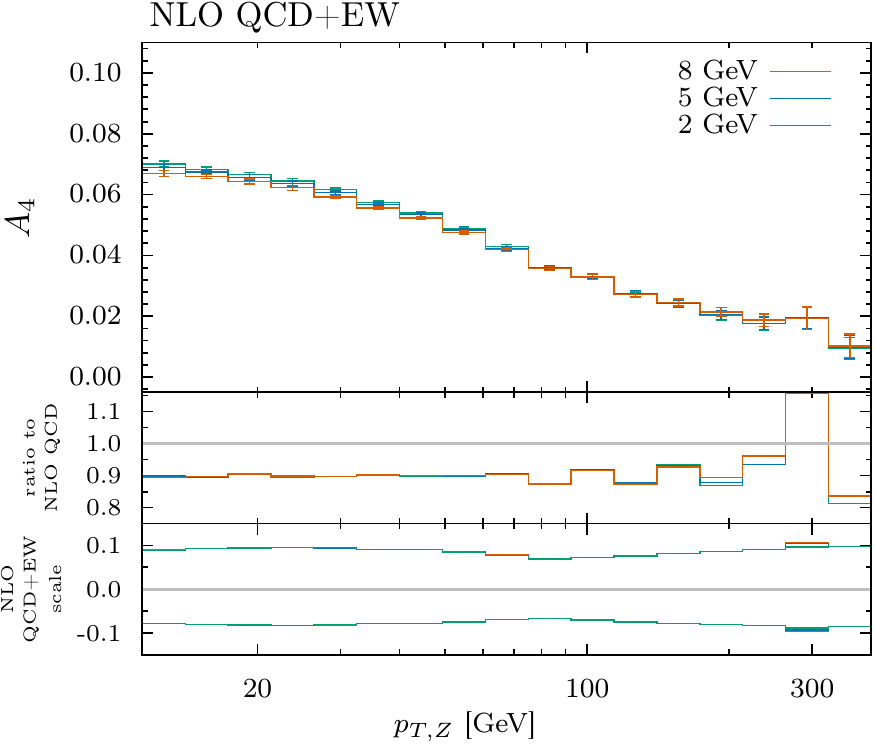}\par 
    \end{multicols}
\begin{multicols}{2}
    \includegraphics[width=0.9\linewidth]{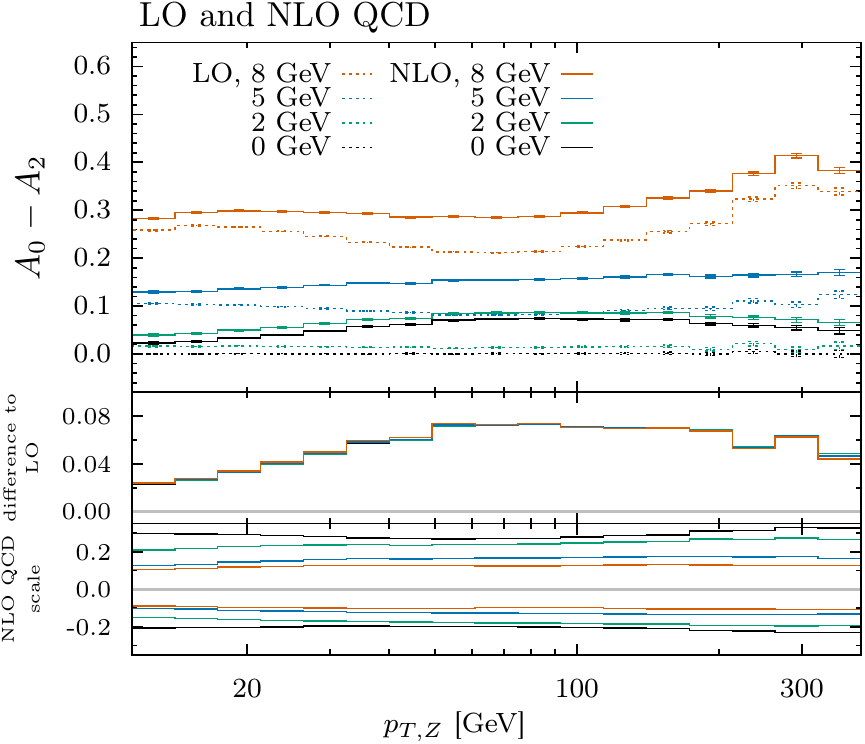}\par 
    \includegraphics[width=0.9\linewidth]{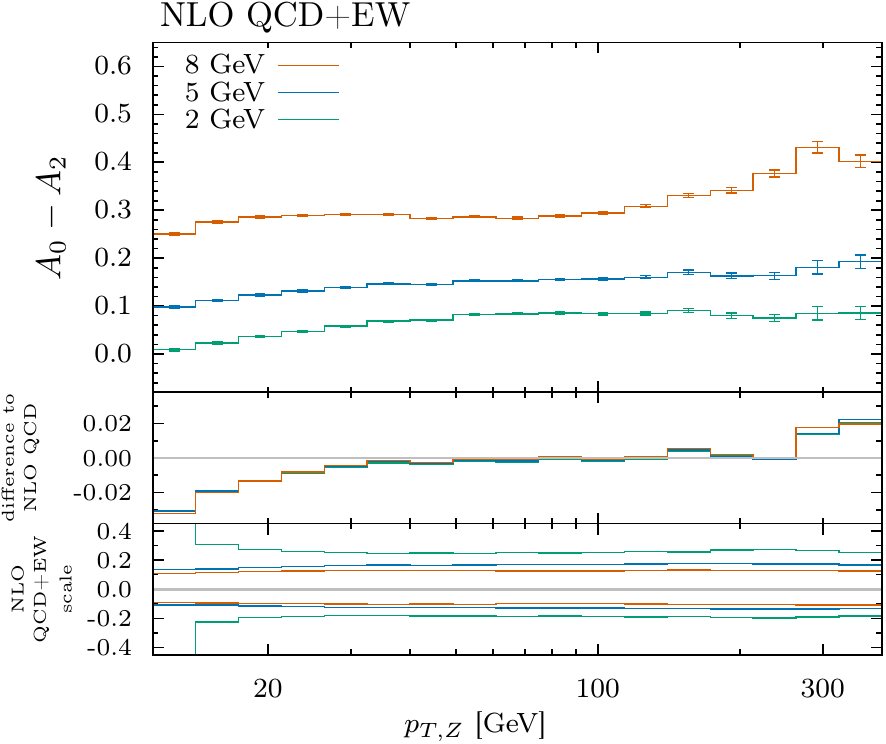}\par
\end{multicols}
\caption{The angular coefficient $A_4$ (top) and the Lam-Tung
  difference $A_0-A_2$ (bottom), the LO and NLO QCD (left) and the NLO
  QCD+EW contribution (right) with corresponding ratios and scale
  uncertainties. The four (left) and three (right) different curves
  indicate the result with different values of the single lepton $p_T$
  cut. See text for more details.}
\label{fig:A4_LT}
\end{figure*}

In Figs.~\ref{fig:A0_A2}-\ref{fig:A4_LT}, we present the angular
coefficients differentially as a function of the lepton pair $p_T$,
following the setup as described in Sec.~\ref{section:setup}. The
layout of all figures is identical. The same angular coefficient
is shown in the left and right plots. In the main panel in the left
plots, the LO (dotted) and NLO QCD (solid) predictions are shown, for
the four values of the single lepton $p_T$ cuts. In particular, the
black, green, blue and red curves correspond to no $p_T$ cut,
$p_T>2.0$~GeV, $p_T>5.0$~GeV, and $p_T>8.0$~GeV, respectively. In the
middle panel, the ratio between the NLO QCD results, over the LO
results are shown, and in the lower panel the uncertainties from scale
variation are displayed for the NLO QCD result. In the figures on the
right, the main panel shows the NLO QCD+EW predictions, now for three
values of the single lepton transverse momentum\footnote{The inclusive
  results, i.e.~without the single lepton $p_T$ cut, are not
  IR-finite, as discussed in Sec.~\ref{section:theory}, and are
  therefore not shown.}. The middle panel shows the ratio of the NLO
QCD+EW predictions over the NLO QCD ones, and the lower panel displays
the scale uncertainties at the NLO QCD+EW level. The only exception to
this layout is in the two plots for the $A_0-A_2$ (lower plots
of Fig.~\ref{fig:A4_LT}) where in the middle insets the difference
between the orders is taken, rather than their ratio.  For all the
coefficients shown here, the LO and NLO QCD predictions (without the
single lepton $p_T$ cut) are in agreement with the corresponding
results presented in Ref.~\cite{Gauld:2017tww}. The results were
checked against distributions obtained for the photon recombination
parameter $R<0.4$ instead of $R<0.1$, and the difference in the final
results is negligibly small for all observables considered here.

In the top two plots of Fig.~\ref{fig:A0_A2} the $A_0$ coefficient is
shown as a function of the transverse momentum of the lepton
pair. From the main panels it is clear that the results, at LO, NLO
QCD and NLO QCD+EW depend significantly on the size of the single
lepton $p_T$ cut. With the cut, the dependence of the coefficient on
the lepton pair transverse momentum is reduced, resulting in a much
flatter distribution. However, as can be seen from the middle panels,
the ratios of the NLO QCD results over the LO ones (left plot) and the
NLO QCD+EW over the NLO QCD ones (right plot), the NLO corrections are
almost completely independent of the single lepton $p_T$ cut, apart
from the region $p_{T,Z}\lesssim 30$~GeV for the NLO EW
corrections. In the latter region, the size of the NLO correction
depends on the single lepton cut, with the smaller the cut, the larger
the NLO correction. This is expected, since this contribution would be
perturbatively unstable when letting the value of the cut go to
zero. For the region $p_{T,Z}\gtrsim 30$~GeV, the overall size of the
corrections is small ---both at the NLO QCD and NLO QCD+EW level it
does not reach more than a couple of percent. In particular for the
size of the EW corrections this is reassuring: since the dependence on
the single lepton $p_T$ cut is negligibly small, one can assume that
the higher order EW effects are also negligible for the predictions
without the single lepton $p_T$ cut. Since the (N)NLO QCD predictions
are in good agreement with the data for this
observable~\cite{Gauld:2017tww}, this remains true with the EW
corrections included as well. The lower inset shows the scale
uncertainties of the NLO QCD (left plot) and NLO QCD+EW (right plot)
predictions. Since the EW corrections are small for this observable,
also the two scale uncertainty bands are of very similar size, which
is about $\pm10\%$ at small and large $p_{T,Z}$ and a couple of
percent points smaller for intermediate $p_{T,Z}$ values.

The $A_2$ coefficient is plotted in the lower two figures of
Fig.~\ref{fig:A0_A2}. Similarly to the $A_0$ coefficient, also for
this coefficient the single lepton transverse momentum cut flattens
the value of the coefficient as a function of the lepton pair
transverse momentum, albeit not in the same way. For this coefficient,
the effect of the cut is much more pronounced at large values of
$p_{T,Z}$, reducing the predictions for this coefficient by up to 50\%
at $p_{T,Z}\approx 300$~GeV. Interestingly, from the two middle panels
one can conclude that the NLO (QCD and QCD+EW) corrections to this
coefficient are almost completely insensitive to the single lepton
transverse momentum cut. Hence, we can safely assume that the relative
contributions from the EW corrections for the inclusive predictions
would be similar in size as to what is given in the middle inset of
the right plot. Indeed, from this middle panel we can see that the EW
corrections are negligibly small for $p_{T,Z}\gtrsim 30$~GeV, but
increase significantly below this value. In particular, they increase
the NLO QCD results by more than a factor two for the smallest $p_{T,Z}$
values shown here. Comparing these corrections to the NNLO predictions
from Ref.~\cite{Gauld:2017tww}, we conclude that the NLO EW
corrections are significantly larger than NNLO, and might overshoot in
the comparison to the data in the first bin somewhat. The size of the
uncertainties estimated from scale variations is similar for the NLO
QCD and NLO QCD+EW predictions.

In the top two plots of Fig~\ref{fig:A1_A3} the $A_1$ coefficient is
presented. As can be seen from the main and middle panels in the left
figure, the NLO QCD corrections enhance the coefficient by up to
$30\%$ at the smallest $Z$-boson transverse momenta probed, but
falling down to close to zero corrections at the largest
transverse momenta ($p_{T,Z}\gtrsim 200$~GeV). These corrections are
almost completely independent from the value of the single lepton
$p_T$ cut. On the other hand, the NLO QCD+EW corrections (as compared
to the NLO QCD corrections alone) are completely flat in this
observable, see the top and middle panels of the figure on the right
hand side. Also these corrections are independent from the single
lepton $p_T$ cut, and it can therefore be assumed that the findings
here can be extrapolated to the inclusive region. Since the EW
corrections are small, the uncertainty from scale variations (lowest
panels in both plots) is not affected to a significant extent by them.

The $A_3$ coefficient is shown in the lowest two figures of
Fig.~\ref{fig:A1_A3}. The QCD corrections are small ($\lesssim 10\%$)
and almost independent from the single lepton $p_T$ cut, as can be
seen from the top and middle panels of the left plot. The NLO EW
corrections on top of the NLO QCD ones are of order
of ten percent throughout the $p_{T,Z}$ interval, and
also here independent of the single lepton cut. Similarly to the other
coefficients, the relative uncertainties from scale variation is
similar for the NLO QCD and NLO QCD+EW predictions (see the bottom
panels of both plots). We remind the reader that in
our LO and NLO QCD predictions we include the dominant EW corrections
to the $\rho$ parameter, see Sec.~\ref{section:setup}. Having included it, the EW correction on top of the QCD correction is somewhat reduced to roughly -10\%. We note however that this is an overall, transverse momentum independent shift, one which is also present in the $A_4$ coefficient (see upper figures in Fig. ~\ref{fig:A4_LT}). As these coefficients are the most sensitive to the weak mixing angle, this overall shift may be a consequence of missing higher order corrections in this parameter. More precisely, the two-loop contribution to the $\rho$-parameter in Eq.~\ref{eq:rho} may mitigate this overall shift in these coefficients. As such,  these overall large electroweak corrections to these coefficients are not an artifact of the perturbative behavior.

The predictions for the $A_4$ coefficient is shown in the top two
plots of Fig.~\ref{fig:A4_LT}. As can be seen from the top panels in
both figures, the predictions for this coefficient are rather
independent from the value of the single lepton transverse momentum
cut. Moreover, also the NLO QCD and EW corrections are independent
from this cut, as can be seen from the two middle panels. The size of
the NLO QCD corrections (of about a few percent) is somewhat smaller than the EW corrections on top of the QCD corrections. Since this
correction is independent from the single lepton $p_T$ cut this can be
extrapolated to the inclusive result.  We point out once more the issue with the sensitivity on the weak mixing angle of this coefficient, as in the case of the $A_3$ coefficient, yielding the large order ten percent EW correction. Would we not have included the one-loop corrections to the $\rho$-parameter, the NLO EW corrections to this coefficient would have resulted in a very large of about $-30\%$ correction over the whole $p_{T,Z}$ range The uncertainty band is not altered significantly after including EW
corrections, which can be seen in the lower panel.

In the bottom two plots of Fig.~\ref{fig:A4_LT} we show the
predictions for the violation of the Lam-Tung relation, i.e.~the left
hand side of Eq.~(\ref{eq:LT}). As expected, see Fig.~\ref{fig:A0_A2},
the dependence on the single lepton transverse momentum cut is
significant for the violation. However, the actual size of the NLO
corrections (both QCD and EW), is rather independent from this cut, as
can be seen from the two middle panels\footnote{We remind the reader
  that these middle insets is not a ratio, but rather the difference
  between the NLO QCD and LO predictions (left plot) and NLO QCD+EW
  and NLO QCD ones (right plot).}. From the middle inset of the right
hand figure, we can conclude that the NLO EW corrections change the
violation of the Lam-Tung relation by up to $-0.03$ at the smallest
$p_{T,Z}$ considered, but reducing with increasing $p_{T,Z}$ and
already compatible with zero at $p_{T,Z}\approx 25$~GeV. A correction
of $-0.03$ is rather significant, since the complete NNLO predictions
in this $p_T$ range are around 0.05 or below~\cite{Gauld:2017tww} when
not imposing the single lepton $p_T$ cut.

\section{Conclusions and discussions}\label{section:conclusions}
In this paper we have examined the five dominant angular coefficients
parametrizing the cross section of $Z$-boson production at
finite-$p_T$ and decay to the leptonic final state $l^+l^-$ at
$\mathcal{O}(\alpha^3\alpha_S)$ at $pp$ collisions at $\sqrt{s}=8$
TeV. We have presented the results differentially in the lepton pair
transverse momentum $p_{T,Z}$. We compared our NLO QCD+EW results to
the NLO QCD predictions, in the invariant mass range $m_{ll} \in
[80,100]$ GeV. We examined the effect of a single lepton transverse
momentum cut, which is included to avoid IR singularities in the
electroweak corrections.

For the variation of the single lepton $p_T$ cut, the general feature
we find is that the coefficients depend on the cut in a similar manner
at all orders of interest: $A_1$, $A_3$ and $A_4$ are found to be the
least affected by the value of the cut at LO, NLO QCD and NLO QCD+EW,
whereas $A_0$ and $A_2$ show a rather significant dependence on this
cut in certain regions. The relative sizes of the dependencies
manifest themselves in the level of the Lam-Tung violation, for which
we find an increasing violation for an increasing single lepton cut,
as expected.  On the other hand, by examining the ratios of
NLO QCD/LO and NLO QCD+EW/QCD it can be concluded that most of the dependence on the single lepton $p_T$ cut factors from the corrections: the
dependence on the cut is similar for the LO, NLO QCD and NLO
QCD+EW predictions. The only minor exception to this can be found for
the low-$p_T$ bins of the $A_0$ coefficient, where an increase in the
cut decreases the significance of the EW corrections (which is
compatible with what can be expected from perturbation theory). Even
though this effect is significant for this coefficient, this does not
affect the Lam-Tung violation, where there is again only a negligible
dependence on the single lepton cut in the size of the NLO QCD and NLO
QCD+EW corrections.

The fact that the relative dependence on the single lepton $p_T$ cut
is almost always negligible for the corrections, allows us to make the
following conclusions about the importance of the fixed-order EW
corrections. For the $A_0$ coefficient the EW corrections are
negligible, except in the region $p_{T,Z}\lesssim 30$~GeV, for which
these predictions cannot be trusted. Similarly, for the $A_2$
coefficient the EW corrections are small, except in the region
$p_{T,Z}\lesssim 30$~GeV, where they rise steeply, resulting in
corrections of more than a factor two on top of the NLO QCD ones. For
the $A_1$ coefficient, the EW corrections are negligible. For the two remaining coefficients $A_3$ and $A_4$, which are sensitive to the weak mixing angle, with our inputs to the calculation, we find a large ten percent EW correction. This, however, is an overall shift, and a even more fine-tuning of this parameter may reduce the size of these corrections.  For the violation of the Lam-Tung relation, the NLO QCD+EW corrections are marginal compared to the NLO QCD corrections for $p_{T,Z}\gtrsim 25$~GeV. However, for
$p_{T,Z}\lesssim 25$~GeV the EW corrections increase, resulting in a
rather significant $-0.03$ correction to the Lam-Tung violation at the
smallest $p_{T,Z}$ values considered.

\begin{acknowledgements}
This work is supported by the Swedish Research Council under contract
number 2016-05996.
\end{acknowledgements}


\bibliography{Bibliography}{}

\end{document}